# Explainable machine learning for breakdown prediction in high gradient rf cavities

Christoph Obermair[*]
*CERN, CH-1211 Geneva, Switzerland and Graz University of Technology, AT-8010 Graz, Austria*

Thomas Cartier-Michaud, Andrea Apollonio, William Millar,[†] Lukas Felsberger,
Lorenz Fischl,[‡] Holger Severin Bovbjerg,[§] Daniel Wollmann, Walter Wuensch,
Nuria Catalan-Lasheras, and Marçà Boronat
*CERN, CH-1211 Geneva, Switzerland*

Franz Pernkopf
*Graz University of Technology, Graz, Austria*

Graeme Burt
*Cockcroft Institute, Lancaster University, Lancaster, United Kingdom*

The occurrence of vacuum arcs or radio frequency (rf) breakdowns is one of the most prevalent factors limiting the high-gradient performance of normal conducting rf cavities in particle accelerators. In this paper, we search for the existence of previously unrecognized features related to the incidence of rf breakdowns by applying a machine learning strategy to high-gradient cavity data from CERN's test stand for the Compact Linear Collider (CLIC). By interpreting the parameters of the learned models with explainable artificial intelligence (AI), we reverse-engineer physical properties for deriving fast, reliable, and simple rule–based models. Based on 6 months of historical data and dedicated experiments, our models show fractions of data with a high influence on the occurrence of breakdowns. Specifically, it is shown that the field emitted current following an initial breakdown is closely related to the probability of another breakdown occurring shortly thereafter. Results also indicate that the cavity pressure should be monitored with increased temporal resolution in future experiments, to further explore the vacuum activity associated with breakdowns.

## I. INTRODUCTION

In the field of particle accelerators, specially designed metallic chambers known as radio-frequency (rf) cavities are commonly employed to establish electromagnetic fields capable of accelerating traversing particles. The energy gain provided by a cavity is determined by the accelerating gradient, a quantity defined as the longitudinal voltage experienced by a fully relativistic traversing particle normalized to the cavity length. Hence, in linear accelerators (LINACS), any increase in the accelerating gradient translates to a reduced machine length. The continued interest in future colliders and other accelerator applications, where machine size is a key constraint, has continued to drive research in this area. One such example is CERN's Compact LInear Collider (CLIC) project, a proposed future high-energy physics facility that aims to collide positrons and electrons at an energy of 3 TeV. To reach this energy at an acceptable site length and at an affordable cost, the project proposes the use of X-band normal-conducting copper cavities operating at an accelerating gradient of 100 MV/m [1].

One of the primary limits on the achievable accelerating gradient in normal conducting high-gradient cavities is a phenomenon known as vacuum arcing or breakdown [2]. To operate reliably at high accelerating gradients, such cavities must first be subjected to a so-called *conditioning period* in which the input power is increased gradually while monitoring for breakdowns [3–5]. Due to the limited understanding of the origin of rf breakdowns and the inability to predict them, current operational algorithms generally act responsively rather than preemptively. Hence, they aim for a progressive recovery of operating conditions by temporarily limiting the rf power following breakdowns [6]. In this paper, we investigate the possibility of employing predictive methods based on machine learning to limit the impact of breakdowns.

---

[*]christoph.obermair@cern.ch
[†]Also at Cockcroft Institute, Lancaster University, Lancaster, United Kingdom.
[‡]Also at Vienna University of Technology, Vienna, Austria.
[§]Also at Aalborg University, Aalborg, Denmark.

Data-driven machine learning algorithms have been successfully deployed in particle accelerator applications for incorporating sequential dynamics using large amounts of available experimental data. Ongoing efforts at CERN have demonstrated the successful use of machine learning for failure analysis in particle accelerators, e.g., to identify and detect anomalies in the rf power source output of LINAC4 [7] or to detect faulty beam position monitors in the LHC [8]. Deep neural networks were used to obtain predictions [9] and its uncertainties [10] in diagnostics for measuring beam properties at SLAC National Lab. At the University of Florida in Gainesville, relevant physical parameters for calculating the critical temperature of new superconducting magnets were discovered [11] with machine learning. Furthermore, eight different superconducting rf faults were classified with high accuracy at Jefferson Laboratory [12] using classic machine learning. However, to the best of our knowledge, none of the stated methods analyzed the parameters of the trained machine learning models, i.e., used explainable-AI, to explore the physical properties of the underlying phenomena. This is particularly relevant when making predictions that have a potential impact on machine protection and machine availability.

Overall, the objective of this work is to (1) analyze historical data of CLIC rf cavities with explainable-AI to better understand the behavior of breakdowns and to (2) investigate possibilities of data-driven algorithms for conditioning and operation of rf cavities.

The paper is organized as follows: Following this Introduction, Sec. II describes the experimental setup and data sources. Section III describes the methodology for data-driven modeling and gives insights into the design choices made, based on the characteristics of the available historical data. We further provide a comprehensive overview of rf-cavity breakdowns, convolutional neural networks for time series, and explainable-AI techniques. We then present the modeling and experimental results for two different data types, i.e., trend data in Sec. IV and event data in Sec. V. With explainable AI, we state that a pressure rise is the first sign of a breakdown and validate it empirically. The strengths and the limitations of our methodology are discussed, together with an outlook for possible future work in Sec. VI. Finally, we conclude our research in Sec. VII.

The code of our machine learning framework is publicly available.[1]

---

[1]https://github.com/cobermai/rfstudies.

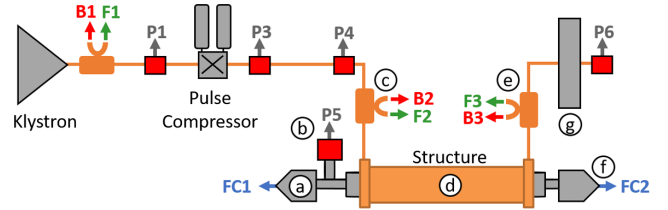

FIG. 1. Schematic of CERN's XBOX2 test stand. The red and green arrows show where the backward reflected traveling wave (B) and the forward traveling wave (F) rf signals are measured via directional couplers. The upstream and downstream Faraday cup signals are labeled FC1 and FC2. The locations of the ion pumps throughout the system are also shown (P). The lowercase letters mark the items also shown in Fig. 2.

## II. EXPERIMENTAL SETUP

To investigate the challenges associated with the high-gradient operation and to validate the novel 12-GHz rf components for the CLIC project, CERN has commissioned three X-band klystron-based test stands named XBOX1, XBOX2, and XBOX3, respectively [13]. The test stands have been previously reported in detail [4,13]. To allow for better readability of this paper, we provide a short introduction to their structure and operation modes. While all three test stands are built with the same arrangement, they mainly vary depending on the specific components used. A schematic of the high-power portion of the XBOX2 test stand is shown in Fig. 1. The locations, denoted with lowercase letters, are also shown in a photograph of one of the test stands in Fig. 2. In each test stand, a 12-GHz phase-modulated low-level radio frequency (LLRF) signal is amplified to the kilowatt level and used to drive a klystron. The high-power rf signal produced by the klystron is then directed through a waveguide network to the rf cavity. To increase the peak power capability, each test stand is also equipped with specially designed energy storage cavities, also known as pulse compressors [14,15].

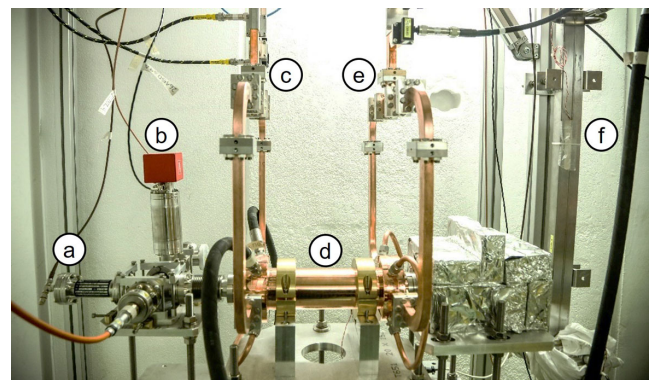

FIG. 2. Picture of a prototype accelerating structure installed in one of the test stands [16]. Visible are the upstream Faraday cup (a), an ion pump (b), the rf input (c) and output (e), the rf cavity under test (d), the shielded lead enclosure (f), and the high-power rf load (g).

During operation, the forward (F) and backward (B) traveling rf signals are monitored via directional couplers. The gradient throughout the waveguide network is measured by directional couplers and logged by the control system. The XBOX2 and XBOX3 test stands are situated in a facility without beam capability. However, during high-field operation, electrons are emitted from the cavity surface and accelerated. This phenomenon, which is undesired in real operation, is known as dark current [17–19]. Monitoring the emitted current during operation is an important measure used in detecting cavity breakdowns, as will be shown later. During the operation of the test stand, the dark current is measured via two Faraday cups, situated on the structure extremities in the upstream (FC1) and the downstream (FC2) directions. Finally, the internal pressure is maintained and measured with a series of ion pumps (P) located throughout the waveguide network.

In Fig. 2, a prototype of the CLIC accelerating structure (d) is visible with the waveguide input (c) and output (e). The directional couplers and coaxial cables, which measure the high-power rf signals, can be seen at the top center, above these waveguide parts. The upstream Faraday cup (a), an ion pump (b), and the high-power rf load (g) are also visible. The downstream Faraday cup is situated inside a shielded lead enclosure (f) which is necessary for protection against the dark current.

Figure 3 shows two examples of different events, measured by the directional couplers and the Faraday cups. On the left side, the data from a healthy event are shown, and on the right side, a breakdown event is plotted. Figure 3(a) shows the approximately rectangular klystron pulse (F1). As is visible in Fig. 1, the test slot is equipped with a pulse compressor. To operate this device, phase modulation is applied to the klystron pulse, beginning after approximately 1700 samples of F1. Note that the position of the edge is not always at the exact position, as it can be changed by the operator without changing the performance of the system. Figure 3(b) shows the resulting "compressed" pulse which is delivered to the structure (F2). The device consists of two narrowband energy storage cavities linked via a hybrid coupler. As a consequence, upon receipt of the klystron pulse, most of the power is initially reflected, resulting in the sharp edge visible after approximately 200 samples (0.125 $\mu$s) of F2. As the storage cavities slowly begin to fill with energy and emit a wave, interference between the reflected and emitted waves occurs, resulting in the gradual change of amplitude in the transmitted waveform. When the phase of the incoming klystron pulse is modulated after approximately 1700 samples (1.0625 $\mu$s) of F2, the reflected and emitted waves constructively interfere, producing a short, high-power region that is flat in amplitude. Following the cessation of the klystron pulse, the remaining energy in the cavities is emitted, resulting in a gradual decay in the amplitude of the transmitted waveform. Further details on the design and operation of the pulse compressor are available in [20].

The signal which is reflected from the structure (B2) is shown in Fig. 3(c). As the accelerating structures are of the traveling wave design, nominally, the reflected signal is small. During breakdown events, however, the arc effectively acts as a short circuit, reflecting the incoming wave as shown on the right of Fig. 3(c). Fig. 3(d) shows the transmitted signal (F3). During normal pulses, this waveform is similar to the signal at the structure's input, while truncation is observed during breakdown events as most of the power is reflected back toward the input [see on the right of Fig. 3(d)]. Finally, the upstream and downstream Faraday cup signals are shown in Figs. 3(e) and 3(f), respectively.

All XBOX2 data are shown in Fig. 4. Specifically, the maximal value and the pulse width of the F2 signal with respect to the cumulative pulses for all data in 2018 are shown. Additionally, the cumulative breakdown count is shown. Initially, many breakdowns occur during the first part of the conditioning. Here, both the F2 maximal value and the pulse width value vary. The yellow area represents pulses, during which these F2 values were stable. These pulses will be used for further processing in Sec. III A.

### A. rf cavity breakdowns

In high-gradient rf cavities, small surface deformations can cause a local enhancement of the surface electric field, resulting in substantial field emission and occasional plasma formation, i.e., arcing, which can damage the surface as shown in Fig. 5. The plasma which forms in the cavity during such breakdown events constitutes a significant impedance mismatch that reflects the incoming rf power.

Additionally, breakdowns are accompanied by a burst of current, which is generally a reliable indicator for structure breakdowns [18,22,23]. Minor fluctuations, which do not lead to the formation of plasma and the subsequent reflection of the incoming power detected by the Faraday cups, are defined as *activity* on the surface of the structure. In the XBOX test stands, these are measured by Faraday cups to reliably detect breakdowns and regulate the conditioning process (see Fig. 2 FC1 and FC2) [3,24]. Typically, at an accelerating gradient of 100 MV/m, Faraday cup signals of the order of 1 mA are observed in the test stands [18]. The threshold for structure breakdowns is typically set to 81.3% of the maximal resolution of the analog to digital converter in the Faraday cups, e.g., −0.615 to 0.615 V for XBOX2, which corresponds to currents in the hundreds of milliamps range. In Fig. 3, it is shown that during breakdown events, a large dark current is emitted, and thus the threshold on the Faraday cup signal

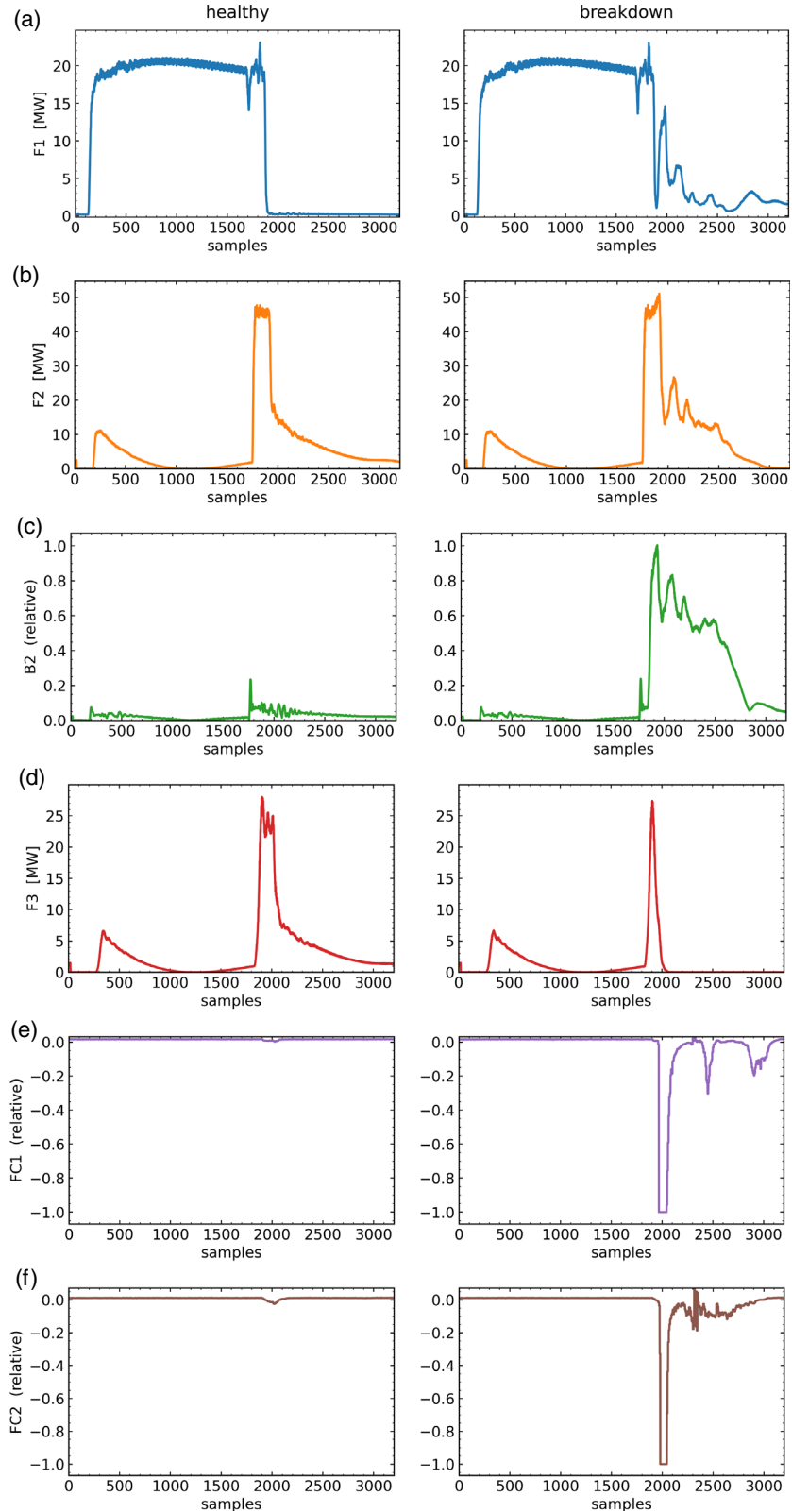

FIG. 3. Two examples of different events, showing the F1, F2, B2, F3, FC1, and FC2 signals. The left plots represent the signals of a healthy event, the right plots represent the signals of a breakdown event. All signals are 2 $\mu$s long. Note that the power amplitude of the forward traveling waves after the klystron (a), before the structure (b), and after the structure (d), are shown in MW. The power amplitude in the backward traveling wave (c), the upstream (e), and downstream (f) Faraday cup signals are shown relative to their maximum value, as no calibration coefficients were provided by the system.

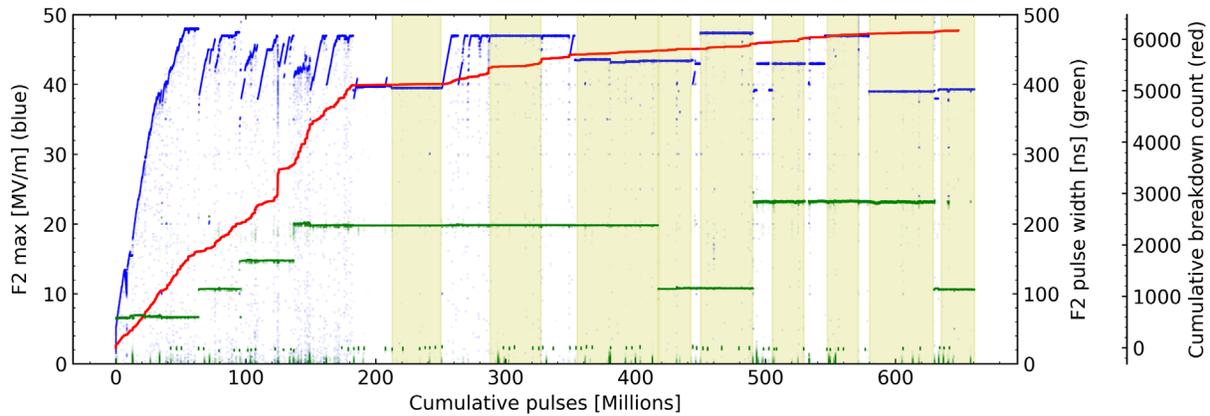

FIG. 4. Overview of the conditioning period, containing all data analyzed. The yellow area represents the runs during which the operational settings were kept stable and which we used for analysis. Additionally, the maximum power amplitude of the forward traveling wave signal F2 (blue), its pulse width (green), and the cumulative breakdown count (red) is shown.

(FC1, FC2) is well suited to distinguishing between healthy and breakdown signals.

Breakdowns usually occur in groups. When a breakdown is detected in the XBOX test stand, the operation is stopped for a few seconds. Afterward, operation is resumed by ramping up the input power within less than a minute.

During conditioning, the total number of breakdowns varies widely on the tested structure, which is why structures are generally more comparable in terms of the cumulative number of rf pulses. As a result, it has previously been proposed that conditioning proceeds primarily on the number of pulses and not solely on breakdowns [25]. This also aligns with the results of high-voltage dc electrode tests, where conditioning has been linked to a process of microstructural hardening caused by the stress associated with the applied electric field [26]. In addition to the copper hardness, the total number of accrued breakdowns is thought to be affected by the copper purity, the cleanliness of the structure [27] defined by the amount of dust and other contamination, the design of the cavity, and the level to which the cavity must be conditioned dependent on the nominal operating power and pulse length.

### B. Data from experimental setup

90 GB of data from a period of 6 months in 2018 were produced during the operation of the XBOX2 test stand. The high-gradient cavity, tested during this time, was produced at the Paul Scherrer Institute in Switzerland [16,28]. The data are divided into so-called *trend* data and *event* data. Trend data contain 30 single scalar values, e.g., pressure measurements, temperature measurements, and other system relevant features. Event data contain six time-series signals of 2 $\mu$s length, with up to 3200 samples (see Fig. 3).

Figure 6 shows an example of the trend and event data logging mechanism. In the test stand, event data are acquired every pulse at 50 Hz and trend data are acquired at up to 600 Hz. Due to the limited data storage of the experimental setup, the data cannot be stored with full resolution. The waveforms associated with an rf pulse are stored in an event data file every minute. In the case of breakdown events, the two prior rf pulses are logged in addition to the pulse, where the breakdown appeared. The corresponding trend data file is updated at a fixed rate every 1.5 s.

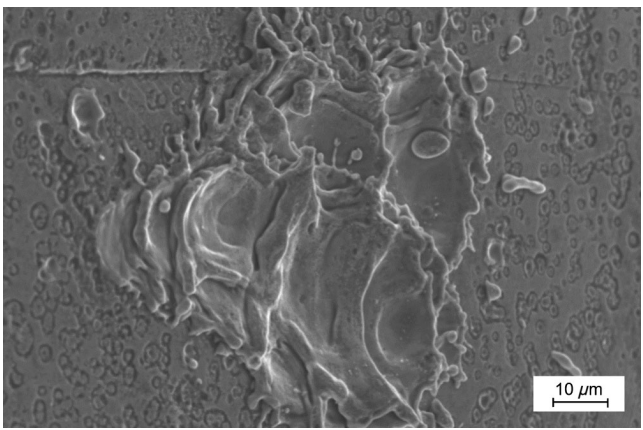

FIG. 5. Example of a crater after a breakdown on the surface of a copper rf cavity [21].

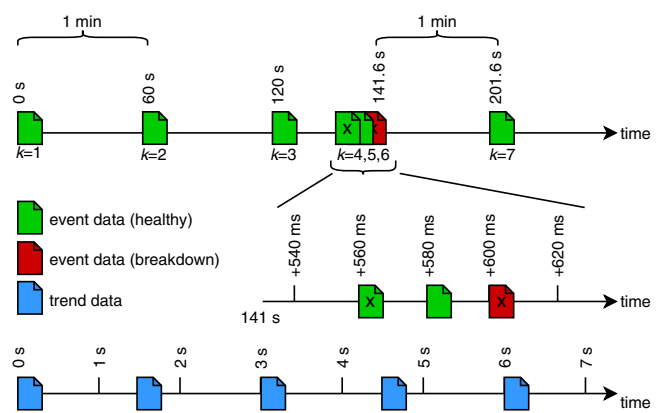

FIG. 6. Trend and event data logging. Event data signals are stored every minute during normal operation. In addition, in case a breakdown occurs, the breakdown event and the two prior healthy events are stored. Trend data features are stored every 1.5 s. The events indicated with an x are not considered for the prediction of breakdowns.

TABLE I. Information about different runs during which the operational setting was stable. Due to the limited amount of breakdowns, groups with similar F2 pulse width are formed for validation and testing during the modeling phase.

| Run | No. of primary breakdowns | No. of follow-up breakdowns | F2 max (MV/m) | F2 pulse width (ns) | Group |
|---|---|---|---|---|---|
| 1 | 10 | 3 | 35.8 | 182.4 | Group 1 |
| 2 | 50 | 58 | 39.5 | 171.2 | Group 2 |
| 3 | 41 | 38 | 34.6 | 161.5 | Test |
| 4 | 14 | 15 | 42.5 | 106.5 | Group 3 |
| 5 | 35 | 62 | 42.7 | 100.8 | Group 4 |
| 6 | 30 | 53 | 38.3 | 211.2 | Group 5 |
| 7 | 21 | 16 | 37 | 186.1 | Group 1 |
| 8 | 13 | 8 | 37.1 | 222 | Group 5 |
| 9 | 5 | 7 | 34.9 | 102 | Group 3 |

To go into more detail on the exact use of machine learning, we describe our data mathematically. Our data are a list of $K$-, $M$-dimensional multivariate time-series $\mathbf{X}_k = [\mathbf{x}_1, \ldots, \mathbf{x}_M]$ for $k \in \{1, \ldots, K\}$. Each of the $M$ time-series has $N$ samples, i.e., $\mathbf{x}_m \in \mathbb{R}^N$ for $m \in \{1, \ldots, M\}$. For both the event and the trend data, an event $K$ is defined as an entry in the event data. The number of time-series $M$ is given by the available signals of the power amplitude of the traveling waves and the Faraday cups for the event data. In the trend data, $M$ is given by the number of available features, e.g., pressure, temperature, and other system relevant features. The number of samples $N$ is defined by the number of samples in the event data signals and the amount of most recent data entries, of an event $k$ in the trend data features.

Based on the Faraday cup threshold stated before, we assign a label *healthy* ($y_k = 1$) and *breakdown* ($y_k = 0$) to each event $k$. This results in a XBOX2 data set of shape $\{\mathbf{X}_k, y_k\}_{k=1}^K$. Using this notation, 124,505 healthy and 479 breakdown events were derived. We further define the first breakdown in each breakdown group as a *primary breakdown*, and all other breakdowns, within less than a minute of the previous breakdown, as *follow-up breakdowns*. With this definition, we split the given 479 breakdowns into 229 primary breakdowns and 250 follow-up breakdowns (see Table I). Compared to the high amount of healthy events, there is only a small amount of breakdown events. This so-called *class imbalance* is tackled by randomly sampling a subset of healthy events and by assigning class weights to the breakdown events during optimization of the algorithm and during the computation of the performance measure.

### III. METHODOLOGY OF ANALYSIS

In this section, we discuss the background of the data processing used to generate the results. Generally, modeling schemes, for representing a system's behavior, are divided into model-driven approaches, where prior knowledge is embedded to represent a system's behavior, and data-driven approaches, where the system's behavior is derived from historical data. With the increasing amount of computational resources, available historical data, and successfully implemented machine learning algorithms, data-driven methods have become popular in many applications for failure prediction [29–31]. The choice of a data-driven algorithm is dependent on the application, the system complexity, and the amount of system knowledge available, as schematically shown in Fig. 7. The goal is to find the simplest model, which is capable to capture the relevant characteristics of the system under study [32].

When considering the goal of identifying a breakdown in an rf cavity, the most common approach relies on an expert setting a threshold [18] on a relevant quantity, e.g., the current measured by a Faraday cup, based on their knowledge about the system. An alternative approach could consider thresholds based on a statistical approach, which can be derived from the distribution of cavity breakdowns from past reliability studies [22]. However, such thresholds are not sufficient for highly nonlinear problems and complex system dependencies, like predicting rf breakdowns. In these cases, classical machine learning models, e.g., k-nearest neighbors (k-NN) [33], random forest [34], and support vector machine (SVM) [35], can be used to find these correlations and to derive optimal, more complex decision boundaries. In k-NN, an event is classified based on the majority class of its neighbors. Here, the neighbors are determined by finding the events with the closest Euclidean distance. A random forest is a combination of many decision trees to an ensemble. Decision trees learn simple decision rules, e.g., the FC1 signal reaches its saturation value, inferred from the most relevant characteristics of the problem, also called features. SVM on the other hand, learns a decision boundary that splits data into classes while maximizing the decision boundary margin. If features in the data are not known *a priori*, deep learning [36], e.g., multilayer perceptrons, or convolutional neural networks, provides the ability to automatically extract and estimate them. Those methods are explained in detail in the modeling subsection. Deep learning can be categorized into discriminative deep learning, which directly models the output based on the input data, and generative deep learning, which models the distribution of the data from which the output is inferred. In order to develop an end-to-end time-series analysis framework without the necessity of manual feature calculations, we use deep learning models to analyze breakdowns in the CLIC rf cavities and show that they achieve superior results compared to classic machine learning approaches, such as k-NN, random forest, and SVM.

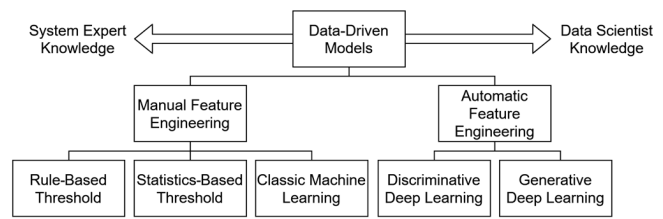

FIG. 7. Overview of different data-driven models.

Specifically, we use discriminative deep learning models, due to their recent success to classify time-series signals [37]. By analyzing our models after training, we show how to extract system knowledge and physics insights, which then allows the extraction of models with reduced complexity.

For the labeled measurement data from the XBOX2 test stand, dedicated python toolboxes are used for feature calculation [38], time-series classification [37], and interpretation of model predictions [39]. Four steps of data processing and analysis, namely, transformation, exploration, modeling, and explanation, are carried out. These are detailed in the next paragraphs.

### A. Transformation

Before training our machine learning models, we apply the following transformation steps to the data. All these steps contribute to fit the data and their properties to our models and include merging of event and trend data, filtering of unwanted events, and resampling and scaling of the event data signals.

*Merging*: Merging and synchronizing the trend data with the event data is a critical data transformation step to ensure the correct relative time order of the data (see Fig. 6). Particular caution is required to take the nearest past trend data samples for each event $k$.

*Filtering*: During our analysis, we only consider data during which the operational setting was stable, i.e., we filter the phases of commissioning or parameter adjustment. Specifically, we define so-called *runs* as the periods where the F2 max and F2 pulse width were kept constant. Table I shows the properties of the different runs, and Fig. 4 highlights these time periods in yellow. Due to the limited amount of breakdowns in certain runs and in order to increase the statistics, we also combine runs with a similar F2 pulse width (see Fig. 3) which we will use for modeling later on. Additionally, using a threshold of 650 kW on the power amplitude of the forward traveling wave signal F2, we further discard all events which only included noise, logged when the machine was actually not operating.

*Scaling*: The used features and signals have different units and different value ranges. To make them comparable, we standardize the data by subtracting the mean and dividing by the standard deviation. This way, all features and signals have a mean equal to 0 and a standard deviation equal to 1, independently of their units.

*Resampling*: In the event data, the Faraday cup signals (FC1, FC2) only have 500 samples compared to the 3200 samples from the other signals, as they are sampled with a lower frequency. Therefore, we interpolate the Faraday cup signals linearly to 1600 samples and selected only every second sample of the other signals.

### B. Exploration

The goal of the exploration phase is to get an initial understanding of the event and trend data and to validate the transformation step. We compute 2D representations of the high dimensional data, in which each data point represents data of an event $k$, e.g., compressing all information that can be found in Fig. 6 on a 2D plane. This enables us to see correlations and clusters within the derived representations in a single visualization of the data. Outlier events, which are fundamentally different from other events, are further analyzed and, if applicable, neglected after further consultation with experts. Representation learning is a key field in machine learning with many methods available including but not limited to unsupervised machine learning methods like principal component analysis [40], stochastic neighbor embeddings [41], and representation learning methods based on neural networks [41–43].

In Fig. 8, we use two dimensional t-distributed stochastic neighbor embedding (2D-tSNE) [41], which converts pairs of data events to joint probabilities, i.e., the likelihood that they are similar. Close events have a high joint probability, and events far away have a low joint probability. Accordingly, 2D-tSNE creates representations in a 2D space and iteratively updates its location, such that the distributions $P$ of the high-dimensional and the 2D space $Q$ are similar. This equals the minimization of the Kullback-Leibler divergence [44] which measures the similarity between two distributions, i.e., $D_{\mathrm{KL}} = \sum_{x \in \mathcal{X}} (P||Q) = P(x) \log(\frac{P(x)}{Q(x)})$, where $\mathcal{X}$ is the domain of $x$.

After the dimension reduction, the different coloring of the representations is used to validate the steps of the transformation phase. No information about the coloring is given to the algorithm during training, which means that neither the runs nor the labels are used as input to compute the 2D-tSNE representations.

Figure 8 shows the 2D-tSNE dimension-reduced representation of the trend data during runs in which the operational settings were kept constant. The axis of the figure represents the two dimensions of the lower dimensional space, where correlations between the data samples are visible. First, representations are automatically colored, identifying the stable runs (a). This leads to clear clusters and validates the separation into different runs. In addition, two clusters with a mix of every run are formed. Their meaning becomes clear with different color schemes. The first cluster with mixed runs gets clear when using a coloring scheme as a result of the filtering in the transformation step (b), i.e., the filtering with the threshold on the power amplitude of the forward traveling wave signal F2.

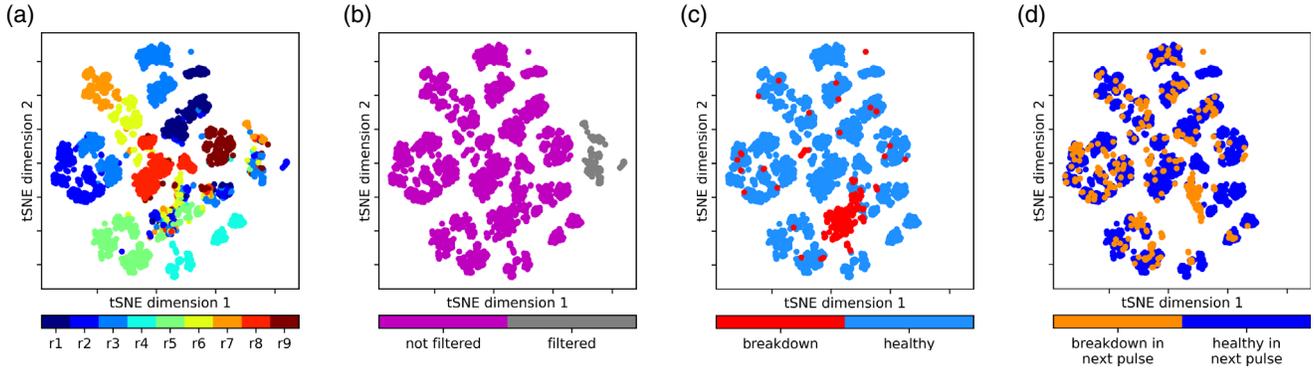

FIG. 8. 2D-tSNE of XBOX2 trend data during stable operation. The algorithm is able to distinguish between (a) stable runs, (b) not filtered and filtered events, and (c) breakdown and healthy events. In (d), no clear separation of events with a breakdown in the next pulse and a healthy event in the next pulse is possible. All representations in (c) are a subset of not filtered events in (b) and all representations in (d) are a subset of all healthy signals in (c).

Using all nonfiltered events from (b), we analyze if it is possible to classify breakdowns without giving the model any information about the label, i.e., if supervised modeling is necessary or if unsupervised learning would already be sufficient. Inspecting the clustering between breakdown and healthy events (c), it seems possible to use unsupervised learning for the classification, as many breakdown events form one cluster and are clearly separable from healthy events. This also explains one of the clusters of signals with mixed runs in (a).

As the unsupervised classification of breakdowns was successful, further investigations aim at identifying breakdowns during the following pulse, i.e., predicting breakdowns. Using all healthy events from (c), no clear unsupervised separation is possible for distinguishing events that are healthy in the next pulse from events that lead to a breakdown in the next pulse (d). Notably, the same phenomena can be observed when using other unsupervised methods, like autoencoders [42] or a higher dimensional space for clustering. As labels are available from the FC signals, we employ supervised learning techniques to distinguish the events shown in Fig. 8(d).

### C. Modeling

The objective of the modeling phase is to find a function $f(X_k)$ that predicts the output $\hat{y}_{k+1}$. This means that we classify whether a *breakdown in the next pulse* $\hat{y}_{k+1}$ will occur. This would be sufficient to protect the cavity and employ reactive measures to prevent its occurrence. The function $f(X_k)$ is modeled with a neural network, and its parameters are optimized during training with the available historical data.

The results are obtained by discarding the event of the breakdown and the event two pulses before a breakdown, expressed with an x in the events $k = 4, 6$ in Fig. 6. This can be attributed to the fact that the equidistance of the event data is violated around a breakdown, which is corrected by this action. The network then solely focuses on using $X_{k=5}$ to predict $y_{k=6}$.

#### 1. Introduction to neural networks

To better understand the behavior of a neural network, we next give a brief overview of its structure. At a single neuron, a weight $w_{m,n}$ is assigned to each input $x_{m,n}$ of $X_k := (x_{0,0}, \ldots, x_{M,N})$. The sum of the input multiplied by the weights is called the activation $a$ of a neuron, which is further used as an input to an activation function $h(\cdot)$. This leads to the following equation:

$$f(X_k) = h\left(\sum_m^M \sum_n^N w_{m,n} x_{m,n} + w_0\right), \quad (1)$$

where $w_0$ is a bias weight. Common activation functions are the sigmoid activation function $h(a) = 1/(1 + e^{-a})$ or the Rectified Linear Unit (RELU) $h(a) = \max(0, a)$. The choice of activation function depends on several factors [36], e.g., the speed of convergence and the difficulty to compute the derivative during weight optimization.

A neural network consists of several layers, where each layer includes several neurons which take the output of the previous layer neurons as an input. This allows the modeling of nonlinear properties in the data set. With a fully connected neural network, a neuron takes all outputs of the previous layer as an input, while in a convolutional neural network (CNN), the neuron only takes neighboring neurons' output of the previous layer as an input. A CNN, therefore, creates correlations with neighboring inputs.

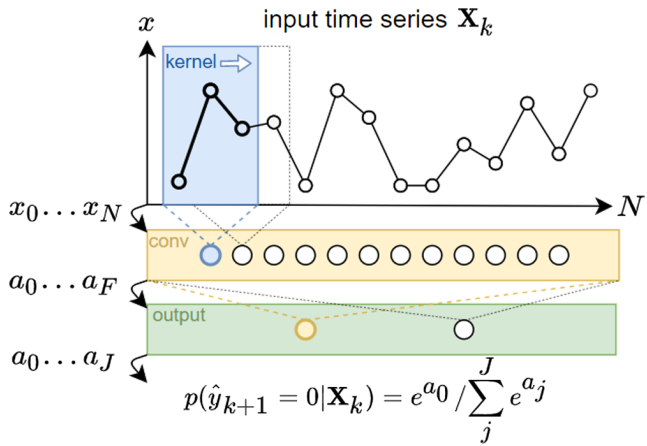

FIG. 9. Example of a convolutional neural network (CNN) for time-series prediction. For simplicity, the input $X_k$ consists of only one signal, i.e., $m = 1$, and the network consists of only one hidden convolutional (conv) layer. As in most of our models, the softmax activation function is used as an output to derive $f(X_k) = p(\hat{y}_{k+1}|X_k)$ out of the activations $a_j$. In this example, the kernel size of the convolution layer is 3, the filter size is $F = 12$, and the probability of a breakdown in the next pulse ($y_{k+1} = 0$), is stated. In this case, the network would have 60 trainable weights.

Essential parameters of a CNN are shown in a simple example in Fig. 9. The *kernel size*, defines the number of neighboring neurons used from the previous layer, and the *filter size*, defines the number of neurons in the current layer. The name filter is derived from the fact that a convolution can also be seen as a sliding filter over the input data. Furthermore, *pooling* refers to the method used for downsampling a convolution to enhance the created correlations. Pooling can be either *local*, over each dimension separately, or *global*, over all dimensions. Two common pooling methods are *maximum pooling*, where the maximum of a window is taken as an output, and *average pooling*, where the mean of a window is taken as an output.

#### 2. Learning of neural networks

Weight optimization is typically achieved with *gradient descent* methods using a loss function. For classification tasks with two classes, typically the cross-entropy-loss $E = -[y \log(p) + (1 - y) \log(1 - p)]$ is chosen, where $y$ is the known class and $p$ is the predicted class probability. In a process with $i$ iterations, called *epochs*, a neuron's weight $w_{m,n}$ is then optimized by $w_{m,n}^{i+1} = w_{m,n}^i - \eta \nabla_w E$. Here, $\eta > 0$ is the learning rate, and $\nabla_w E$ is the gradient of the loss dependent on the weights. The gradient descent optimization can be further accelerated with more sophisticated optimizers. Specifically, we use the ADAM optimizer [45] in our models. It varies the learning rate dependent on the mean and the variance of the gradient. In Fig. 14(b), we visualize the learning process of our models, by showing the models' loss with respect to the epochs.

#### 3. Advanced architectures

Due to their ability to learn correlations of neighboring inputs, CNNs contributed to the recent success of machine learning, finding many applications in image classifications [46], language processing [47], and time-series classification [37].

(i) *time-CNN*: The time CNN was originally proposed by Zhao *et al.* [48] and consists of two average pooling convolutional layers with 6 and 12 filters with a kernel of size 7. It uses the mean-squared error instead of the categorical cross-entropy-loss [44] for weight optimization, which is typically used in classification problems. Consequently, the output layer is a fully connected layer with a sigmoid activation function. Due to this architecture, the time-CNN has 4976 trainable weights and is therefore the model with the fewest parameters in our studies.

(ii) *FCN*: The fully convolutional network was originally proposed by Zhao *et al.* [49] and consists of three convolutional layers with 128, 256, and 128 filters of kernel size 8, 5, and 3. In each layer, batch normalization is applied, normalizing the output of the previous layer in each iteration of the weight optimization [50]. This leads to faster and more stable training. Each convolutional layer uses a RELU activation function, except the last one, where the output $a_1, \ldots, a_J$ is globally averaged and fed into a softmax activation function $h_i(a_1, \ldots, a_J) = e^{a_i}/\sum_j^J e^{a_j}$ to obtain the output probability $p(\hat{y}_{k+1}|X_k)$ for $i = 1, \ldots, J$, where $J$ is the number of different labels. The model has 271,106 trainable weights.

(iii) *FCN-dropout*: It is of similar architecture as the FCN with the same number of 271,106 trainable weights. In addition, it has two dropout layers after the second convolution and the global average pooling layers as proposed by Felsberger *et al.* [29]. This dropout layer is skipping neurons during training randomly with a probability of $p_{\text{drop}} = 0.5$, which improves the generalization of the model.

(iv) *Inception*: Inspired by the Inception-v4 network [51], an inception network for time-series classification has been developed [52]. The network consists of six different inception modules stacked to each other, leading to 434,498 trainable weights. Each inception model consists of a so-called *bottleneck layer*, which uses a sliding filter to reduce dimensionality and therefore avoids overfitting. Additionally, several filters are slided simultaneously over the same input and a maximum-pooling operation is combined with a bottleneck layer to make the model less prone to small perturbations.

(v) *ResNet*: The residual network was originally proposed by Zhao *et al.* [49] and consists of three residual blocks, i.e., a group of three convolutional layers. This architecture leads to 509,698 trainable weights. This relatively deep architecture is enabled by using skip connections after each block. This skip connection is a shortcut over the whole block and provides an alternative path during weight optimization which reduces the risk of vanishing gradients [36]. The kernel size of the convolutional layers is set to 8, 5, and 3 in each residual block for the fixed number of 64 filters in each layer. The activation function, the batch normalization, and the output layers are similar to the FCN.

All models were trained on a single Nvidia Tesla V100 GPU. This took on average 24 min for the event data and 9 min for the trend data. Once the models were trained, one prediction took 27 ms for the event data and 18 ms for the trend data using TensorFlow [53] to compile the model without any optimization or compression. However, due to the random weight initialization and depending on the network, the training time slightly varied.

When using a softmax activation function in the last layer, the output of the function in Eq. (1) is the probability of the next event being healthy or a breakdown, i.e., $p(\hat{y}_{k+1}|X_k) \in [0,1]$. To receive a binary label, $\hat{y}_{k+1} \in \{0,1\}$, it is necessary to set a threshold to the probability. The receiver operating characteristic (ROC) curve is a plot that shows how this threshold impacts the relative number of correctly classified labels as a function of the relative number of falsely classified labels. The ROC curve of the best models for each prediction task is shown in Fig. 14(a). We use the area under the ROC curve (AR) to rate the performance of our models. This is a measure of the classifier's performance and is often used in data sets with high class imbalance [54]. Intuitively, this score states the probability that a classifier designed for predicting healthy signals ranks a randomly chosen healthy event $k^+$ higher than a randomly chosen breakdown event $k^-$, i.e., $p[f(X_{k^+}) > f(X_{k^-})]$. An AR score of 1 corresponds to the classifier's ability to correctly separate all labels, while an AR score of 0 represents the wrong classification of all labels.

For training, validation, and testing of our model, we merged runs with similar F2 pulse width into groups as shown in Table I, as some runs have a small number of breakdowns. Specifically, we use *leave-one-out-cross-validation* on the groups. This means we iterate over all possible combinations of groups, while always leaving one group out for validation. After adjusting the model weights, e.g., the class weight, we then test our model on data from run 3.

The mean score $AR_\mu$ over all iterations and its standard deviation, $AR_\sigma$, are stated in the results together with the test result $AR_t$. In order to ensure that our model provides a good generalization to new data, we aim that $AR_t$ of the test set should be within $AR_\mu \pm 2AR_\sigma$. To compare deep learning models with classic machine learning models, we additionally present the AR score of k-NN, random forest, and SVM algorithms. The hyperparameters of these models have been optimized during a sensitivity analysis. Specifically, we used $k = 5$ neighbors for k-NN, $t = 500$ decision trees in random forest, and the *radial basis function* for the SVM, with $C = 1, \gamma = 3.3 \times 10^{-2}$ for trend data and $C = 1, \gamma = 7.2 \times 10^{-5}$ for event data. For a detailed description of these hyperparameters, we refer to existing literature [44].

### D. Explainable AI

To interpret the "black box" nature of deep neural networks, explainable AI recently gained attention in domains where a detailed understanding of the driving factors behind the results is of primary importance. In fields like medical applications [55,56], criminal justice [57], text analytics [58], particle accelerators [29], and other fields in the industry [59], experts cannot simply accept automatically generated predictions and are often even legally obliged to state the reasons for their decision. To reliably predict breakdowns in rf cavities, the explanation of a model is of similar importance. Hence, we utilize explainable AI in our studies to provide the experts with any relevant information used by the model to aid in interpreting the behavior of data-driven models, build trust in the prediction, validate the results, and find possible errors within the earlier data processing steps. Additionally, understanding why a prediction is made may shed light on the underlying physics of vacuum arcs and thus aid in future design decisions pertaining to high-gradient facilities.

Explainable AI is divided into event-wise explanation, where each prediction of the model is analyzed separately, and population-wise explanation, where all predictions are investigated at once. Event-wise explanation enables experts to gain trust in a specific prediction. The choice of event-wise explanation algorithms is dependent on the input, i.e., image, text, audio, or sensory data, and the preferred explanation technique, i.e., by showing the sample-importance [60] or by explanation-by-example [61]. Important samples are often computed with additive feature attribution methods [60,62,63], which calculate a local linear model for a given event to estimate the contribution of a feature to one prediction. Alternative gradient-based methods aim to determine the features that triggered the key activations within a model's weights [64,65]. Explanation-by-example states reference examples on which the prediction is made, by using the activation of the last hidden layer in a neural network and searching for similar activations of events in the training set [61].

Population-wise explanation helps experts to gain trust in the model and to select relevant input features for the predictions. In its simplest form, this is achieved with a greedy search [66], or deep feature selection [67] which applies similar techniques to regularized linear models [34,68]. However, both of the stated methods are very computationally intensive for deep learning models. A more efficient method proposes to train an additional selector network to predict the optimal subset of features for the main operator network [69].

In our studies, event-wise explanations are converted into population-wise explanations by looking at the distribution of a subset of event-wise explanations [70]. Our event-wise explanations are calculated with an additive feature attribution method [60]. This means we define a model

$$g(X_k) = \sum_{m}^{M} \sum_{n}^{N} \phi_{m,n} x_{m,n} + \phi_0, \quad (2)$$

which is approximating the output $f(X_k)$ for one event $k$, where $X_k$ is either the trend data or the event data. In this local linear model, $\phi_{m,n}$ equals the contribution of the feature $x_{m,n}$ to the output $f(X_k)$ and is called the *feature importance*. To calculate $\phi_{m,n}$, we assign a reference value to each neuron. This reference value is based on the average output of the neuron. When a new input value $x_{m,n}$ is fed into the network, a contribution score is assigned to the neuron, based on the difference between the new output and the reference output. All contribution scores are then back-propagated from the output to the input of the model $f$, based on the rules from cooperative game theory [71]. The contribution scores $\phi_{m,n}$ at the input are called SHapley Additive exPlanation (SHAP) values [39] and are used to explain our produced results.

This interpretation is, however, different for trend and event data. In trend data, the SHAP values are interpreted as feature importance, stating the contribution of, e.g., the pressure to the prediction of breakdowns. In event data, the SHAP values are given for each time-series sample, e.g., the importance of each of the 3200 samples in the F1 signal. Here, the mean of all SHAP values in one signal is taken to derive the overall importance of a signal.

## IV. RESULTS USING TREND DATA

In this section, we report the results of applying the methodology of analysis described above, using the trend data of the XBOX2 test stand. Specifically, we use the $N = 3$ closest trend data point in the past, of an event $k$, as described in Sec. II B. Each trend data event consists of $M = 30$ values, including pressure, temperature, and other system relevant features, measured in the test stand.

### A. Modeling

Table II shows the AR score for the prediction of breakdowns with trend data. The results of the different model types described in the previous section are reported for comparison and discussed in detail. For each type of breakdown, the best model score is highlighted in bold. We chose the best model based on four decision criteria: (i) the average performance of the model $AR_\mu$, (ii) the ability of the model to generalize within runs $AR_\mu \pm 2AR_\sigma$, and (iii) the ability of the model to generalize to new data $AR_t$. Additionally, we consider (4) the simplicity of the model given by the number of trainable weights and the complexity of the model structure, as this has a direct impact on the computational cost, which we want to minimize.

The ResNet model is able to predict primary breakdowns with an average AR score of 87.9%. With 7.2%, the standard deviation is much higher compared to the prediction of follow-up breakdowns, but still, the best generalization capability compared to the other models for predicting primary breakdowns. The inception network scores best on the test set with 82.9%. However, since the ResNet model performs best on two out of four decision criteria, we consider it the best for predicting primary breakdowns.

The relatively high standard deviation in the prediction of primary breakdowns states that the patterns learned by the network vary, i.e., the indicators of a primary breakdown differ dependent on the runs on which the network is trained.

With an $AR_\mu$ score of 98.7% and an $AR_t$ score of 98.6%, the inception model predicts follow-up breakdowns best. This means that for the training set, there is a probability of 98.7% that our model assigns a higher breakdown probability to a randomly chosen breakdown event than it assigns to a randomly chosen healthy event. The score is 0.1% less when the model uses the test data. This indicates that the model generalizes well to new data, as the $AR_t$ score is within $AR_\sigma$. The ResNet model offers similar results and an even smaller $AR_\sigma$. However, the inception model is preferred for the prediction of follow-up breakdowns due to its fewer trainable weights.

TABLE II. AR score of different models, predicting primary, follow-up, and all breakdowns with trend data. The model for each column is highlighted in bold. $AR_\mu$ relates to the average AR score of different validation sets and $AR_\sigma$ to the standard deviation. The trained model is finally tested on the test set with a performance $AR_t$.

|  | (1) Primary breakdowns | | | (2) Follow-up breakdowns | | | (3) All breakdowns | | |
|---|---|---|---|---|---|---|---|---|---|
| Model | $AR_\mu$ (%) | $AR_\sigma$ (%) | $AR_t$ (%) | $AR_\mu$ (%) | $AR_\sigma$ (%) | $AR_t$ (%) | $AR_\mu$ (%) | $AR_\sigma$ (%) | $AR_t$ (%) |
| k-NN | 61.0 | 7.4 | 63.1 | 89.8 | 8.1 | 92.9 | 76.7 | 8.0 | 75.9 |
| SVM | 68.8 | 10.0 | 73.8 | 93.6 | 5.7 | 97.0 | 84.2 | 9.8 | 87.8 |
| Random forest | 81.0 | 16.7 | 82.5 | 96.9 | 3.5 | 96.5 | 87.9 | 13.3 | 90.0 |
| Time-CNN | 55.2 | 11.0 | 48.1 | 92.8 | 3.8 | 87.6 | 67.7 | 6.3 | 66.0 |
| FCN | 86.1 | 8.7 | 81.0 | 98.2 | 1.0 | 97.8 | **93.8** | **4.2** | **90.6** |
| FCN-dropout | 84.9 | 9.0 | 81.7 | 95.6 | 3.0 | 97.3 | 92.7 | 4.6 | 90.6 |
| Inception | 85.4 | 8.5 | 82.9 | **98.7** | **1.6** | **98.6** | 92.3 | 4.8 | 90.9 |
| ResNet | **87.9** | **7.2** | 80.4 | 98.7 | 1.4 | 98.0 | 93.1 | 4.6 | 90.1 |

Looking at the prediction of both follow-up and primary breakdowns, the AR scores are approximately averaged compared to the two separate AR scores, the number of primary and follow-up breakdowns is similar. This indicates that the model finds similar patterns for both breakdown types. Here the FCN model scores best with an $AR_\mu$ score of 93.8% and an $AR_\sigma$ of 4.2%. While the $AR_t$ score of 90.6% is slightly lower than in the inception model, the FCN model has significantly fewer trainable weights.

The time-CNN model generally performs poorly compared to the others. A possible reason for this is that the low amount of trainable time-CNN weights cannot capture the complexity of the data. Additionally, the structure of the model might be insufficient. Here, we specifically refer to the unusual choice of Zhao et al. [48] to select the mean-squared error and not the cross-entropy-loss. The mean-squared error is typically used in regression problems, where the distribution of data is assumed to be Gaussian. However, in binary classification problems, the data underlie a Bernoulli distribution, which generally leads to better performance and faster training of models trained with the cross-entropy-loss [72]. The lower performance of the time CNN suggests that the mean-squared error should not be used in classification tasks for XBOX2 breakdown prediction.

Random forest is the only classic machine learning algorithm that achieves similar $AR_\mu$ and $AR_t$ scores compared to deep learning. For example, when looking at the prediction of primary breakdowns, the $AR_t$ score of 82.5% is even higher than the ResNet score of 80.4%. However, the standard deviation $AR_\sigma$ of 16.7% is more than twice as high compared to the ResNet model, which makes its prediction less reliable. The higher standard deviation of classic machine learning compared to deep learning is also observed in the other breakdown prediction tasks.

For each prediction task, the ROC curve of the best model's test set performance is shown in Fig. 14(a). Here, the true positive rate corresponds to the percentage of correctly predicted healthy events, and the false positive rate corresponds to the amount of falsely predicted healthy events. For predicting primary breakdowns, the ResNet ROC curve (1) is plotted in green. Note that the $AR_t$ score, corresponding to the area under the ROC curve, is 80.4% in this case. One can see a slow rise, which reaches a true positive rate of 1.0 at a false positive rate of about 0.4. For predicting follow-up breakdowns, the inception model (2, red) has the highest $AR_t = 98.6\%$ which is confirmed by the large area under the red curve. The curve of the FCN (3, blue) for predicting all breakdowns with $AR_t = 90.6\%$, is a mixture of the primary and follow-up breakdown prediction curves. It is reaching a true positive rate of 1.0 at a false positive rate of about 0.2. Using this information, it can be decided at which probability $p(\hat{y}_{k+1} = 1|X_k)$ an event should be classified as a healthy event. Considering the inception model (2, red) for predicting follow-up breakdowns, a good choice would be the "edge," where the true positive rate is $\sim 1$ and the false positive rate is 0.05. Here, almost all healthy events are labeled correct, while 5% of all breakdowns are falsely considered to be healthy events. However, the final choice of the probability threshold depends on the final application setting of the model and the consequences of false positives and false negatives, further discussed in Sec. VI.

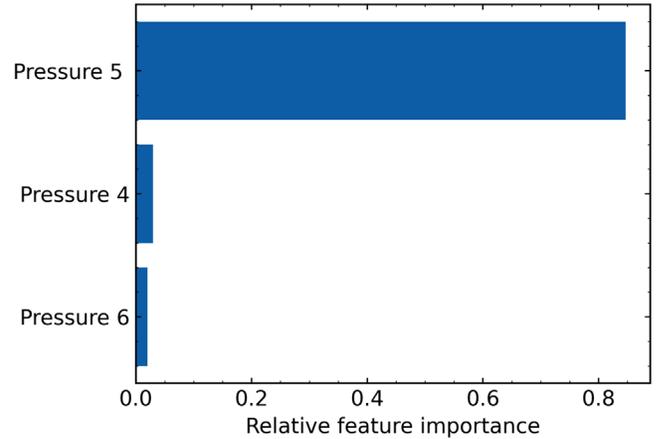

FIG. 10. The three most important trend data features, selected from 30 features in total, for predicting primary breakdowns with trend data.

### B. Explainable AI

As primary breakdowns are generally considered a stochastic process [73], the good performance in Table II on predicting primary breakdowns is especially interesting. Hence, we focus on the trained models to gain deeper insights into the reason behind the good prediction results.

Figure 10 shows the importance of the features $X_k$ for the prediction of primary breakdowns with trend data. Pressure 5 measurements, indicated also with P5 in Fig. 1, is the most relevant feature by a very significant margin, even when compared to the second and third most relevant features. By looking at this signal in more detail, for the different breakdown events in Fig. 11, it can be seen that the highest pressure reading is logged up to a few seconds before a breakdown event. Initially, it was expected that the pressure should be highest after the breakdown is detected via the Faraday cups, after the arc formation and the burst of current. However, here we observe the peak value beforehand.

We investigated the possibility that the observed effect is caused by a systematic error or a timing misalignment in pressure rise, which could have occurred due to the logging algorithm in the control software of the XBOX2 test stand. We utilized a trend data feature of the XBOX2 test stand, which indicates whether the test stand was in an interlocked state, i.e., pulsing is inhibited, or if it is pulsing. Notably, this feature was not used for prediction. Since the pulse rate is 50 Hz, we know that the breakdown must have occurred in 1 of the 75 pulses prior to the interlock. Figure 11 shows the trend data features of the internal beam-upstream pressure during run 4. All data are aligned to the interlock time of the mentioned XBOX2 feature, which is indicated with the black dashed line. The gray area is the confidence interval, covering the previous 75 pulses during which a breakdown occurred, and the interlock signal was generated. A rise in pressure is visible in all data samples before the interlock is triggered. However, the low trend data sampling frequency means significant aliasing is possible, and so the true peak pressure could occur either earlier or later than is shown in the data. Therefore, the internal beam-upstream pressure signal should further be investigated.

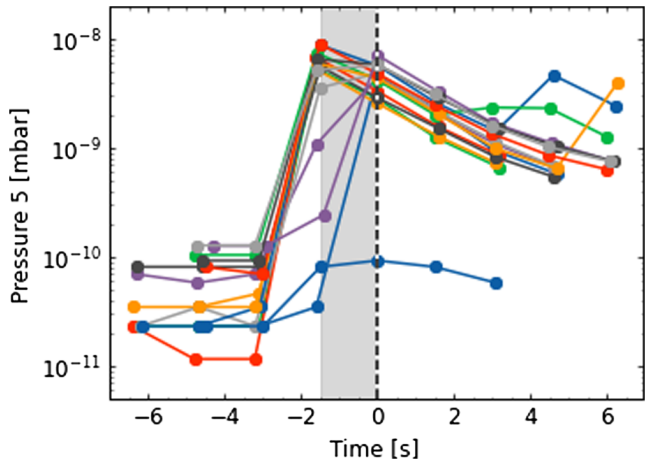

FIG. 11. Data samples of pressure 5, aligned to the interlock state of the test stand. The gray area represents the confidence interval, i.e., the window of time covering the previous 75 pulses in which the breakdown occurred. Data indicate that the pressure begins to rise before an interlock is triggered with the Faraday cup and the reflected traveling wave signals.

Notably, during breakdowns, the vacuum readings located the furthest away from the structure demonstrated a markedly smaller rise which occurred later in time than that observed in the pumps located closest to the structure. This aligns with the expectation that the pumps situated farthest from the site of a given pressure change should measure it last due to the vacuum conductivity of the waveguide.

Generally, significant outgassing is observed in the early stages of component tests in the high-gradient test stands, and a conditioning algorithm that monitors the vacuum level and regulates the power to maintain an approximately constant pressure has been designed specifically for this early phase of testing [13]. It is known, that the exposure of fresh, unconditioned surfaces to high-electric fields results in measurable vacuum activity, however, it is unclear why a measurable pressure rise may occur prior to breakdown when a stable high-gradient operation has been reached. One potential explanation is that the phenomenon may be related to the plastic behavior of metal under high fields. In recent years, it has been proposed that the movement of glissile dislocations, which is a mobile dislocation within the metal, may nucleate breakdowns if they evolve into a surface protrusion [74]. If such dislocations migrate to the surface, then the previously unexposed copper may act as a source for outgassing, resulting in measurable vacuum activity while also being liable to nucleate a breakdown soon thereafter.

### C. Experimental validation

To experimentally validate the phenomenon of the pressure rise before the appearance of a breakdown in the XBOX2 test stand, a dedicated experiment was conducted on a similar rf cavity in the XBOX3 test stand. In case of a substantial pressure increase which may indicate a vacuum leak, klystron operation is inhibited and thus no further high-power rf pulses can be sent to the structure. To facilitate interlocking, the pumps throughout the waveguide network are checked at 600 Hz, several hundred Hz higher than the rf repetition rate. However, due to the limited storage space, not all data are logged (see Fig. 6).

If the pressure begins to rise several pulses prior to a breakdown event, then by appropriately setting the threshold, it is possible to generate an interlock signal and stop pulsing prior to the breakdown. If the rise in pressure is caused by the start of processes that lead to a breakdown then by resetting the interlock and resuming high-field operation, it is assumed that the processes may continue, and a breakdown will then occur shortly after the initial interlock was generated. To validate this hypothesis, a 3-h test slot was granted in CERN's XBOX3 test stand during which the threshold for vacuum interlocks was set to be abnormally low, close to the pressure, at which the test stands generally operate. During this time slot, the data in Fig. 12 was recorded. The procedure of the experiment is visualized in Fig. 13. After detecting the early pressure rise with explainable AI, this finding allows us to simply use a threshold above 10% of the nominal pressure (see Fig. 11). Naturally, a large sample size, i.e., number of primary breakdowns, is desirable to validate the phenomenon. The breakdown rate may be considerably increased by raising the operating gradient although, as shown in Fig. 11, the pressure remains considerably elevated following breakdown events, necessitating a recovery period of several minutes before the pressure returns to the prebreakdown baseline. Additionally, increases in power are associated with increased vacuum activity and so stable, low pressure operation was favored throughout the run to avoid false alarms and ensure reliable interlocking. During the 3-h experiment period, five primary breakdowns occurred, two of which were preceded by a vacuum interlock. One such example is shown in Fig. 12.

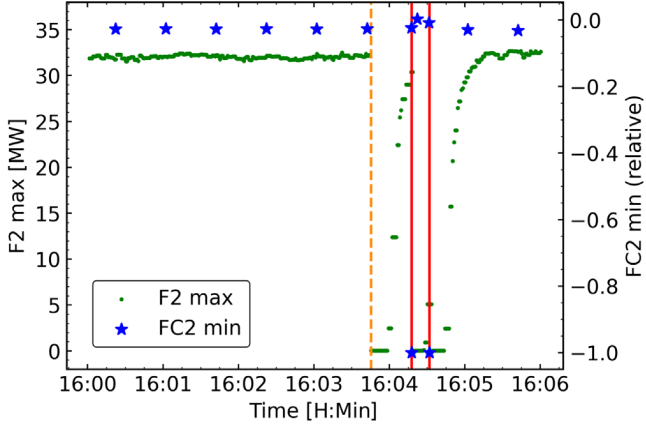

FIG. 12. Maximum value of the structure input power amplitude of the forward traveling wave (F2 max) and minimal value of the downstream Faraday cup signal (FC2 min) during the experiment to predict breakdowns. The orange dashed line shows an interlock, activated by a threshold on the pressure signal, meant to prevent a breakdown. The maximum structure input power amplitude of the forward traveling wave is logged as a feature in the trend data every 1.5 seconds. The minimal value of the downstream Faraday cup signal is extracted from the event data according to Fig. 6.

In Fig. 12, an interlock was produced and then reset several seconds later. The reset was done by removing the interlock thresholds temporarily to allow the test stand to ramp back up to nominal conditions and resume high-power operation. After ramping up in power, two primary breakdowns occurred, as shown by the red lines.

These instances align with what was observed in the historical data. However, given the relatively few primary breakdowns, further experiments are necessary. To overcome the alignment and resolution issues present in the historical data, an improved test stand logging system is currently being developed to record pressure as event data with high resolution.

## V. RESULTS USING EVENT DATA

In this section, we report the results of applying the methodology of the analysis described above, using only the event data of the XBOX2, as shown in Fig. 3. We report these results separately to show that our models do not solely rely on the pressure reading as described in the previous section to successfully predict breakdowns.

### A. Modeling

In Table III, we summarize the results of predicting breakdowns with event data based on the models described in Sec. III. We use the same decision criteria as in the previous Sec. IV A to select the best model.

With a mean validation score of 56.6% and a test score of 54.0%, the FCN-dropout performs best on the prediction of primary breakdowns. Although the $AR_\sigma$ score of 8.3% is higher than in the inception model, the FCN-dropout model is preferred since it has significantly fewer trainable weights. Note that a score of 50% equals a random classifier, which guesses the output. Despite the stochastic behavior of primary breakdowns, our models exceed the expected 50%. However, the result is significantly lower compared to the prediction of primary breakdowns with trend data in Table II. This shows that the pressure rise found in analyzing the trend data is the main indicator for predicting primary breakdowns, given the available data and the described models.

Nevertheless, using event data, the models accurately predict follow-up breakdowns. Here the FCN model is preferred with an AR score of 89.7% for the prediction of follow-up breakdowns and shows the best generalization result on the test set with 91.1%. The AR score of 89.7% implies that with a probability of 89.7%, the FCN model attributes a higher breakdown probability to a randomly selected breakdown event than a randomly selected healthy event. The FCN-dropout offers better generalization on different runs with an $AR_\sigma$ of 5.3%, but relatively bad generalization on the test set with an $AR_t$ score of 8.7%. The inception model and the ResNet model archive similar results, but utilize more trainable weights, which is disadvantageous.

TABLE III. AR score of different models, predicting primary, follow-up, and all breakdowns with event data. The model for each column is highlighted in bold. $AR_\mu$ relates to the average AR score of different validation sets and $AR_\sigma$ to the standard deviation. The trained model is finally tested on the test set with a performance $AR_t$.

| | (4) Primary breakdowns | | | (5) Follow-up breakdowns | | | (6) All breakdowns | | |
|---|---|---|---|---|---|---|---|---|---|
| Model | $AR_\mu$ (%) | $AR_\sigma$ (%) | $AR_t$ (%) | $AR_\mu$ (%) | $AR_\sigma$ (%) | $AR_t$ (%) | $AR_\mu$ (%) | $AR_\sigma$ (%) | $AR_t$ (%) |
| k-NN | 49.6 | 1.2 | 48.4 | 61.4 | 10.1 | 58.7 | 57.2 | 10.0 | 54.9 |
| SVM | 50.0 | 0.0 | 50.0 | 63.0 | 7.8 | 62.5 | 57.3 | 3.6 | 56.3 |
| Random forest | 48.2 | 3.4 | 50.0 | 66.9 | 9.2 | 73.0 | 58.4 | 6.9 | 59.7 |
| time-CNN | 52.7 | 3.4 | 51.9 | 79.2 | 12.8 | 82.1 | 59.8 | 7.7 | 66.6 |
| FCN | 54.7 | 9.8 | 52.8 | **89.7** | **8.1** | **91.1** | 66.8 | 12.5 | 68.7 |
| FCN-dropout | **56.6** | **8.3** | **54.0** | 89.1 | 5.3 | 83.7 | **65.2** | **7.3** | **67.3** |
| Inception | 52.6 | 3.6 | 49.9 | 87.9 | 8.4 | 90.5 | 65.9 | 13.6 | 67.1 |
| ResNet | 51.9 | 7.0 | 53.5 | 88.7 | 7.7 | 89.9 | 67.2 | 14.3 | 68.5 |

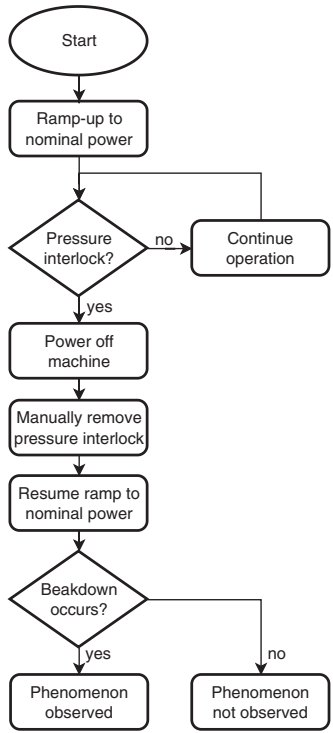

FIG. 13. Flowchart showing the procedure of the experiment. The pressure interlock was set to 10% above a nominal pressure. The Faraday cup signals and the reflected traveling waves were used to detect the breakdown.

With 8.1%, the standard deviation of predicting follow-up breakdowns with event data is much higher than the prediction of follow-up breakdowns with trend data in Table II. This means that the patterns learned by the network vary more when our models are trained on event data than on trend data. The values in Table I underline this conclusion, as the F2 max values and the F2 pulse width values are different depending on the run. The influence of the F2 max deviation is mitigated by the standardization of each signal by its own mean. However, the fluctuation of the F2 pulse width values makes it harder for the network to find common patterns in the time-series signals. In the trend data, the model mainly focused on the pressure rise, which is a phenomenon occurring across all runs.

Like in Table II, the mean of both primary and secondary breakdown prediction scores is close to the prediction of all breakdowns. This again indicates that the patterns detected are used for both follow-up and primary breakdowns. However, in primary breakdowns, this pattern occurs only rarely, leading to lower performance compared to the prediction of breakdowns with trend data. Here, the ResNet model has the best $AR_\mu$ score with 67.2%, the FCN-dropout model has the best $AR_\sigma$ score of 7.3%, and the FCN model has the best $AR_t$ score with 68.7%. Overall, the FCN-dropout model is considered best, due to the significantly lower standard deviation and the relatively low amount of trainable weights compared to the inception model.

In contrast to the trend data results in Table II, all classic machine learning methods show lower performance than the deep learning models. Figure 7 shows that classic machine learning requires features as input. When those features are given, as they are in the trend data, similar performance to deep learning is achieved. However, in the event data, time-series signals are used as input instead of features. Classic machine learning models are not able to generalize well anymore. Deep learning models automatically determine features in their first layers, and therefore, reach higher performance in all three prediction tasks.

Figure 14(a) shows the ROC curve of the best model's test set performance from Tables II and III. For predicting primary breakdowns, the FCN-dropout model (4, cyan) with $AR_t = 54.0\%$ is close to the orange dashed random classifier, where with $AR = 50.0\%$. Contrary, the FCN model (5, purple) for predicting follow-up breakdowns with $AR_t = 91.1\%$ covers a significantly larger area under the curve. The FCN-dropout model (6, black) combines the two curves, indicating that the predicted breakdowns were mostly follow-up breakdowns.

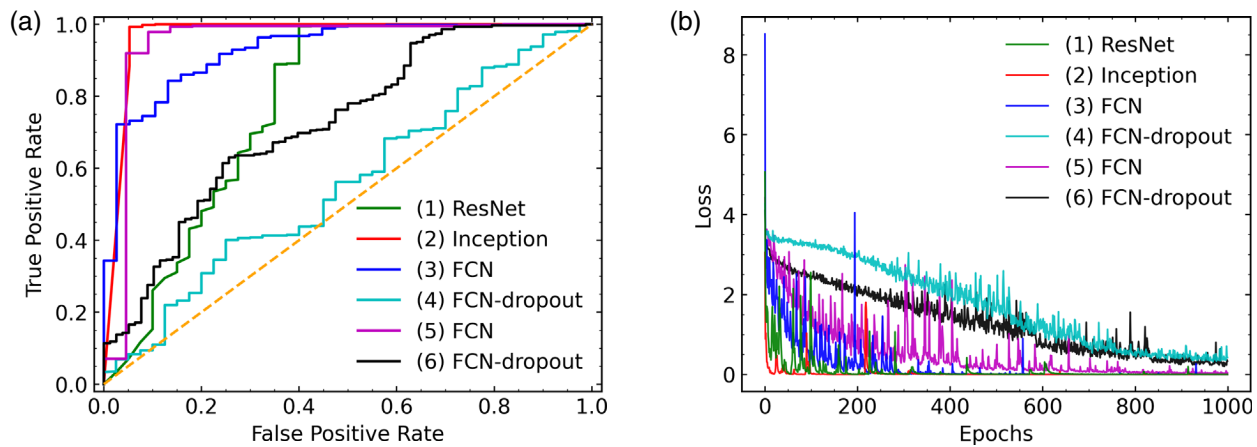

FIG. 14. Receiver operating characteristic (ROC) (a) and learning curve (b) of trend and event data modeling. For all prediction tasks (1–6) shown in the results in Table IV and Table V, the curves of the best model's test set is shown. The dashed orange line represents a random classifier in the ROC curve.

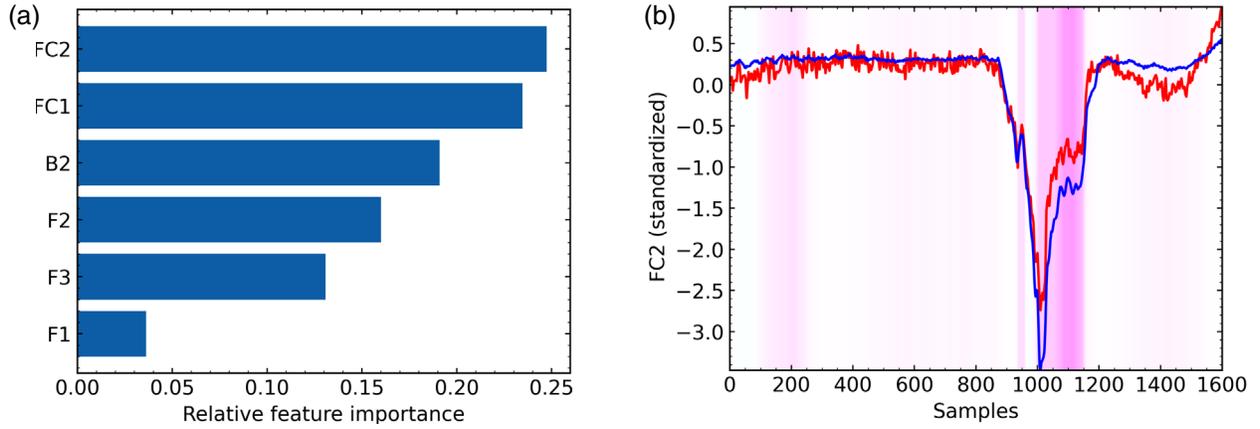

FIG. 15. Most important signals (a) and FC2 samples (b) for predicting follow-up breakdowns with event data. In addition to the most important samples (marked by the pink background), the average preceding signal for a subsequent healthy event (blue) and a breakdown event (red) is shown, respectively.

Similar to the trend data prediction, the threshold on $p(\hat{y}_{k+1} = 1|X_k)$ can be selected. For example, there are two "edges" in the (5, purple) ROC curve at a false positive rate of about 0.05 and at 0.2. At the first "edge," ∼50% of all healthy events are classified correctly, and only 5% of breakdowns are falsely considered healthy. At the second "edge," ∼90% of all healthy events are classified correctly, but 20% of breakdowns are falsely classified as healthy. The selected threshold is dependent on the class weight, as we use $124{,}505 \times 2.5\% \approx 3113$ healthy and 479 breakdown events, and the effect on the machine availability of the application, as discussed in Sec. VI.

However, the number of epochs in our experiments is not fixed. The models are trained until the loss does not change significantly within 100 epochs, i.e., we use early stopping. Figure 14(b) shows the learning curve for the test set prediction of all the best models for 1000 epochs.

Models trained on trend data (1–3) converge faster than models trained on event data (4–6). In addition, models trained on follow-up breakdowns (2,5) converge faster than models trained on primary breakdowns (3,6). Also, the performance of classic machine learning models is closer to deep learning models in follow-up breakdowns compared to primary breakdowns. This indicates that correlations within the data and follow-up breakdowns are more linear compared to correlations within the data and primary breakdowns. The FCN-dropout model (4, cyan) for predicting primary breakdowns and the FCN-dropout model (5, black) fail to converge to a loss close to zero. This is in good agreement with the fact that those models achieve lower $AR_t$ scores.

### B. Explainable AI

Due to the poor performance for the prediction of primary breakdowns, only models for the prediction of follow-up breakdowns are considered for the explanation in this section.

The signals identified by the FCN as being most important for the prediction of follow-up breakdowns are shown in Fig. 15. The downstream Faraday cup signal (FC2) is classified as being most important (a) by the used models, but the difference to the other signals is not as significant as in Fig. 10. Further investigation showed that a specific portion of both Faraday cup signals, particularly the rising edge, was identified by the SHAP approach as being the most important region for breakdown prediction.

An example is shown with the downstream Faraday cup signals in Fig. 15(b). Here, the mean signal over all "healthy in the next pulse" events is plotted in blue and the mean over all "breakdown in the next pulse" events is plotted in red. The important samples in the signal, i.e., the SHAP values, are highlighted in pink. The most important area for the model is approximately 1000–1200 samples.

The reason for a relatively high noise in the red signal is twofold. First, there is higher variance in breakdown signals, as they generally vary in their shape. Second, follow-up breakdowns are generally lower in amplitude. This is due to the fact that after the machine is stopped as a consequence of a primary breakdown, its input power is gradually increased again to recover the nominal power. This leads to lower amplitudes in the follow-up breakdown signals. We mitigate this effect by standardizing each signal separately with its own mean and standard deviation. However, due to the lower amplitudes, the noise is more severe in follow-up breakdown signals. The increased deflection at the end of the red signal is also attributed to this effect. Notably, our models do not focus on the noise or the deflection at the end, because the rising edge of both Faraday cup signals enables more general predictions.

The identified portion in the signal in Fig. 15 has been previously studied in detail [17,22]. The shape of the dark current signal is generally defined by several quantities. The fill time, i.e., the time for the rf pulse to propagate from the input to the output of the prototype CLIC structures, is generally in the order of 60 ns, which corresponds to 48 samples in the plot. As the rf pulse fills the structure of the individual cells, i.e., the subsection in the rf cavity, the cells begin to emit electrons. This results in a rising edge in the F1 signal which is comparable to the fill time of the structure. A similar transient behavior is then observed at the end of the rf pulse, as the structure empties and the cells stop emitting.

Breakdowns alter the surface of the rf cavity and thus change the emission properties of the structure. As a consequence, both the amplitude and shape of the signal are known to change markedly after breakdowns [73,75]. It is postulated that particular signal characteristics may then be associated with an increased probability of future breakdowns. Additionally, it has previously been proposed that fluctuations in the dark current signal may be associated with nascent breakdowns, however, these fluctuations have proven difficult to measure [22]. Such fluctuations constitute another phenomenon that could potentially be detected with the present framework. Notably, all previous observations seem compatible with the findings and explanations of our ML studies.

## VI. FUTURE WORK

The goal of our study is twofold. First, we want to shed light on the physics associated with breakdowns through the insights gained with explainable AI. Second, we aim at supporting the development of data-driven algorithms for conditioning and operation of rf cavities based on machine learning. In this section, we further elaborate on these goals and future activities, starting from the results presented in the previous paragraphs.

### A. Breakdown Physics

To further validate the explainable-AI findings in this work, future experiments will focus on the validation of the presence of a pressure rise prior to the occurrence of breakdowns, by using our simplified threshold-based model to provide an interlock signal. To make more insightful explanations, especially suited for the domain experts of CLIC, we will further improve the used explainable-AI algorithms. Current explainable-AI methods are developed and tested mostly with the goal to interpret images and highlight important areas for classification problems. Typical examples involve the recognition of characteristic features of animals, e.g., the ear of a cat. In images, those areas are self-explanatory and easy to understand by humans. However, explanations in time-series signals are harder to interpret (see Fig. 15). In the future, our work will focus on refining the model explanations by investigating the possibility of using understandable features and correlations to the important areas, e.g., the low mean value and high frequency in the important area of the red signal in Fig. 15. For this, we will build on existing work, which searches for correlations in the activations of the hidden CNN layers [61,76–79].

### B. Model application

Investigations on the direct application of our models are ongoing. Here, the final model will be selected depending on the chosen task according to Tables II and III. For example, the FCN would be chosen for predicting follow-up breakdowns with event data, as it performs best. Below, we address several remaining challenges with which the model's performance could be improved and the potential of machine learning further exploited. Additionally, it is currently under evaluation of how the predictive methods can be embedded in the existing system by notifying an operator or by triggering an interlock before a predicted breakdown.

*Model improvements.*—To further advance the development of data-driven algorithms for conditioning and operation, we will test and improve our model with data from additional experiments. The accuracy of machine learning models is highly dependent on the quality of the data with which the model is trained. As such, the importance of continuous and consistent data logging during experiments is of primary importance during the study and further improvements are being discussed with the CLIC rf test stand team to (i) increase the logging frequency for both trend and event data, (ii) to implement signals of additional pressure sensitive sensors, e.g., vacuum gauges and vibration sensors, or (3) provide a means of accurate timing calibration in the test stand.

*Model embedding.*—As mentioned in Sec. II, it has previously been proposed that accelerating structures condition on the number of cumulative rf pulses and not solely on the cumulative number of breakdowns [25]. This also aligns with the intuition that conditioning is a process of material hardening caused by the stress of the applied electric field [26]. As such, possibilities are investigated to increase the applied field at a rate that still produces the hardening effect but refrains from inducing breakdowns unnecessarily frequently. Conversely, as conditioning typically requires on the order of hundreds of millions of pulses, it is highly desirable to minimize the number of pulses taken to reach high-field operation in order to reduce the electricity consumption and test duration. The optimal method may lie between these two scenarios, where our machine learning models come in to improve future conditioning algorithms.

Second, we focus on the possibility to derive operational algorithms that are planned to increase machine availability in modern high-gradient accelerators, exploiting our machine learning models. The basic idea is to maximize the availability of a future accelerator by dynamically detuning structures that are predicted to experience a breakdown, thus limiting the impact of breakdowns on the operation. The reduction in energy associated with doing so may then be compensated in one of two ways, either by powering an additional, spare structure in the beam line which is normally desynchronized, or alternatively, by temporarily increasing the voltage in the remaining structures until the arcing structure stabilizes again. In this scenario, the effect of false predictions of our model will directly affect the performance of the machine, and it is therefore of crucial importance to achieve sufficient accuracy in the predictions.

In a single rf structure, the approach discussed above is no longer valid. Currently, if a breakdown is detected, it is unclear if the breakdown is inevitable or if it may be avoided by taking an appropriate action. If the implemented response is one which interlocks the machine temporarily, a false prediction would then result in an unnecessary stop of the machine and hence a reduction in availability equal to that associated with the breakdown event. Thus, in such a scenario, a threshold on the probability of $p(\hat{y}_{k+1}|\boldsymbol{X}_k)$ is preferred such that the classification is healthy if the model is uncertain. Alternatively, a hybrid model [80] could be implemented, e.g., to enable machine operators to adjust the machine parameters if there are many predicted future breakdowns.

## VII. CONCLUSION

In the work presented, a general introduction to data-driven machine learning models for breakdown prediction in rf cavities for accelerators was shown. Following the steps of transformation, exploration, modeling, and explanation, several state-of-the-art algorithms have been applied and have proven to be effective for our application. By interpreting the parameters of the developed models with explainable AI, we were able to obtain system-level knowledge, which we used to derive a fast, reliable, and threshold-based model.

We have shown that our models can predict primary breakdowns with 87.9% and follow-up breakdowns with an AR score of 98.7% using trend data. Thanks to the analyses carried out with explainable AI, we discovered that historical CLIC rf test bench data indicate that the pressure in the rf cavity begins to rise prior to the Faraday cup signals, in case of a breakdown. Our findings could enable the possibility to act before a breakdown is detected with the Faraday cup signal by setting a lower threshold on the vacuum signal. This would allow us to either avoid the breakdown development at an early stage or to take additional actions to preserve the beam quality.

Using event data, we achieved an AR score of 56.6% for predicting primary breakdowns and 89.7% on follow-up breakdowns, highlighting the low capabilities of the model to predict primary breakdowns but high performance on follow-up breakdowns. Focusing on the latter, explainable-AI points out that the last part of the rising edge in the Faraday cup signals has a high influence on the occurrence of breakdowns. Investigations to explain this behavior are currently ongoing but are supported by past studies on the subject.

Our code is publicly available[1] and provides a framework for the transformation, exploration, and modeling steps, which can be used to analyze breakdowns in other fields or domains.